\documentclass[pra,aps,floatfix,10pt,twocolumn,nofootinbib,superscriptaddress]{revtex4-1}

\usepackage{upgreek}

\usepackage[T1]{fontenc}
\usepackage[utf8]{inputenc}
\usepackage[english]{babel}
\usepackage{microtype}
\usepackage{graphicx}
\usepackage{amsmath}
\usepackage{amssymb}
\usepackage{amsthm}
\usepackage{mathtools}
\usepackage{dsfont}
\usepackage{booktabs}
\usepackage{bm}
\usepackage{xcolor}
\usepackage[caption=false]{subfig}
\usepackage{placeins}
\usepackage{needspace}

\usepackage{tikz}
\usetikzlibrary{arrows.meta}

\newcommand{\dd}{\mathrm{d}}
\newcommand{\ii}{\mathrm{i}}
\newcommand{\Tr}{\operatorname{Tr}}
\newcommand{\I}{\mathds{1}}
\newcommand{\e}{\mathrm e}
\newcommand{\TV}{\mathrm{TV}}
\newcommand{\ket}[1]{\ensuremath{|#1\rangle}}

\usepackage{quantikz}

\theoremstyle{plain}
\newtheorem{theorem}{Theorem}[section]

\usepackage[colorlinks=true,citecolor=blue,urlcolor=cyan,linkcolor=blue]{hyperref}
\begin{document}
\raggedbottom

\title{\texorpdfstring{Quenched Thermodynamics from Ancilla-Assisted Imaginary-Time Evolution}{Quenched Thermodynamics from Ancilla-Assisted Imaginary-Time Evolution}}

\author{Ibukunoluwa Adisa}
\affiliation{Department of Physics, University of Maryland, College Park, MD 20742, USA}
\affiliation{Joint Quantum Institute, University of Maryland, College Park, MD 20742, USA}

\author{Daniele Cuomo}
\affiliation{Department of Physics, University of Jyv\"askyl\"a, FI-40014 Jyv\"askyl\"a, Finland}

\begin{abstract}
Quenched thermodynamics combines the equilibrium physics of fixed coupling configurations with a prescribed distribution of disorder. {\color{black}Problems related to quenched thermodynamics are generally intractable to approximate. We provide a quantum algorithm to calculate quenched disorder averages for the transverse-field Ising model, whose coupling configurations are organized into gauge classes. We show how to estimate these averages by preparing approximate Gibbs states through ancilla-assisted imaginary-time evolution. At each measurement, the ancilla record identifies the gauge class to which the energy estimate contributes. 

We validate our framework through} an open \(3\times3\) lattice. A one-step circuit resolves all sixteen flux classes, and a finite-shot calculation estimates the quenched energy for independently distributed bond signs. Thermodynamic integration extends the reconstruction to a quenched free-energy estimate, while comparison with exact diagonalization identifies the finite-step approximation as the dominant source of error. This use of the measurement record provides access to the thermodynamics of frustrated spin systems with a disorder distribution chosen independently of the circuit's sampling frequencies.
\end{abstract}

\maketitle

\section{Introduction}
\label{sec:introduction}

Imaginary-time evolution connects quantum-state preparation to equilibrium statistical mechanics by suppressing high-energy components of a state and, when applied to a maximally mixed input, produces a thermal state. For a Hamiltonian \(H\) at temperature \(T\), let \(\beta=1/(k_{\rm B}T)\) denote the inverse temperature, with \(k_{\rm B}\) the Boltzmann constant. The partition function normalizes the Gibbs state and determines its thermal expectation values,
\begin{equation}
\rho_\beta=\frac{\e^{-\beta H}}{\Tr\e^{-\beta H}},
\quad
\langle H\rangle_\beta=\Tr H\rho_\beta.
    \label{eq:thermal-expectation}
\end{equation} 
Implementing this nonunitary evolution on a quantum processor requires an indirect construction. Variational methods approximate its action within a family of accessible states, while measurement-based methods embed the desired operation in a larger circuit containing auxiliary qubits~\cite{Motta2020QITE,McArdle2019VITE,Yuan2019VQS,mao2023measurement,kondappan2023imaginary,Chen2023Thermal}.

In a probabilistic construction, selecting a particular auxiliary outcome realizes a chosen imaginary-time direction. The ancilla-resolved protocol of Ref.~\cite{Cuomo_2026} retains both outcomes of each elementary interaction. The measurement result determines the sign of the corresponding Hamiltonian term. A full measurement record therefore identifies a Hamiltonian with known coupling signs. For an Ising chain with open ends, local spin transformations remove these signs, and every record describes the same thermal physics in a different spin basis. Closing a loop changes the situation because the product of coupling signs around the loop is invariant under those transformations. This product is a \(\mathbb Z_2\) flux, familiar from the description of frustration in signed Ising models and lattice gauge theory~\cite{wegner1971duality,toulouse1987theory,morita2005gauge}.

The surviving fluxes make ancilla records useful for studying quenched disorder. In a quenched model, the couplings remain fixed while the spins equilibrate, and thermodynamic quantities are subsequently averaged over a specified distribution of those couplings. Records with the same fluxes belong to one gauge class and share its thermal properties. Resolving these classes therefore provides the ingredients for such a disorder average. Such disordered and frustrated models are challenging to characterize because the number of classes grows exponentially with system size.
Computing the thermodynamic properties of disordered quantum spin systems can be computationally intractable in general~\cite{Bravyi2022QuantumHamiltonianComplexity}. In particular, approximating the partition function of the transverse-field Ising
model is \texttt{NP}-hard in certain parameter regimes
\cite{sinclair_et_al:LIPIcs.TQC.2026.1}.

We use the measured record to identify a class, estimate its energy from the accompanying spin measurements, and combine the class energies with a chosen disorder distribution. Because that distribution enters at the averaging stage, the same resolved data can describe several disorder models. The circuit's class frequencies determine how well those averages can be sampled, while the prescribed distribution determines which physical model they describe.

We study this reconstruction numerically for the transverse-field Ising model on a square lattice with open boundaries. A \(3\times3\) lattice contains four elementary square loops, whose fluxes distinguish sixteen gauge classes. Simulations with a finite number of measurement shots resolve their energies and reconstruct an average for independently distributed bond signs. Extending the calculation across temperature gives a quenched free-energy estimate through thermodynamic integration. Comparisons with exact diagonalization (ED) identify the error introduced by the one-step imaginary-time circuit and distinguish it from sampling and integration errors. Together, these results establish how information retained in the ancilla record can be used to reconstruct the thermodynamics of a prescribed disordered model.

\begin{figure*}[t]
    \centering
    \includegraphics[width=0.98\textwidth]{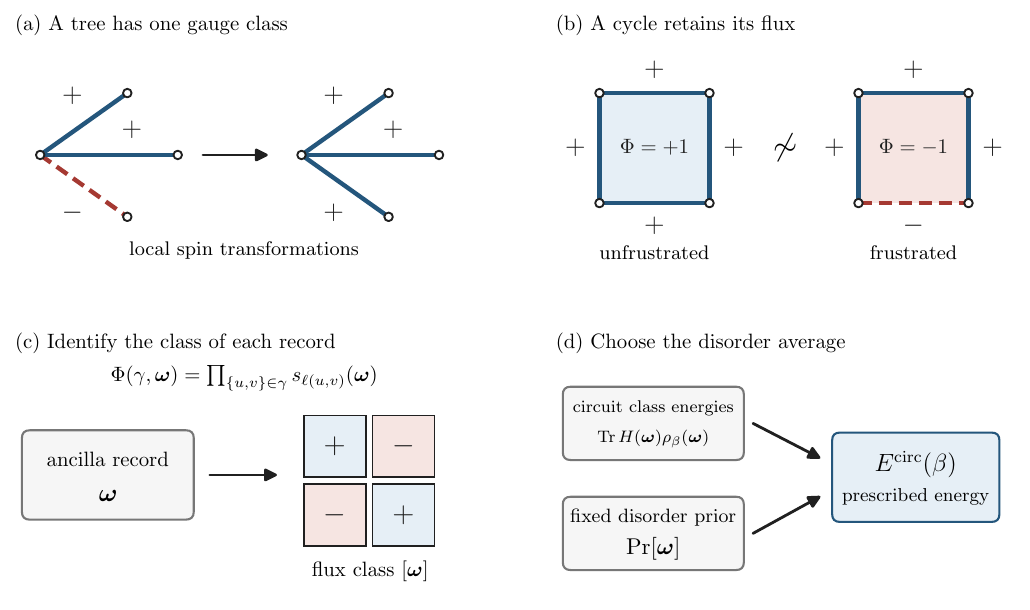}
    \caption{
    Using ancilla records to reconstruct disorder averages.
    (a) On a tree, local spin transformations make all bond signs positive.
    (b) On a square loop, the sign product \(\Phi\) survives these transformations.
    \(\Phi=+1\) and \(-1\) distinguish unfrustrated and frustrated configurations.
    (c) Multiplying the recorded bond signs around each plaquette gives
    \(\Phi(\gamma,\boldsymbol{\omega})\) and identifies the flux class \([\boldsymbol{\omega}]\).
    (d) The class-conditioned circuit energies
    \(\operatorname{Tr}H(\boldsymbol{\omega})\rho_\beta(\boldsymbol{\omega})\), combined with the fixed
    disorder prior \(\Pr[\boldsymbol{\omega}]\), give \(E^{\mathrm{circ}}(\beta)\).
    Finite-shot measurements estimate this quantity as \(E^{\mathrm{shot}}(\beta)\) and
    replacing the circuit energies by Gibbs expectations gives the target
    \(E(\beta)\). Integrating the prescribed circuit average over inverse
    temperature gives the free-energy estimate developed in Sec.~\ref{sec:reconstruction}. 
    }
    \label{fig:workflow}
\end{figure*}

\section{From measurement records to gauge classes}
\label{sec:records}

The measurement record specifies the Hamiltonian associated with a circuit run. Local spin transformations identify records that describe equivalent physics, leaving the cycle fluxes as labels for the distinct classes. For energy measurements, these transformations are accounted for in classical processing of the recorded signs and spin outcomes, as described in Appendix~\ref{app:gauge}. This connection between the measured circuit and the lattice geometry underlies the reconstruction.

\subsection{Conditional imaginary-time evolution}

The elementary circuit follows the ancilla-resolved construction of Ref.~\cite{Cuomo_2026}, coupling one auxiliary qubit to one term of the system Hamiltonian. For a system of \(n\) spins, write the Hamiltonian as a sum of \(m\) Pauli operators \(P_j\) with real coefficients \(c_j\),
\begin{equation}
    H = \sum_{j}^{m}c_j P_j,
    \qquad P_j^2=\I,
    \label{eq:H-pauli}
\end{equation}
where \(\I\) is the identity on the \(2^n\)-dimensional system Hilbert space. For a single Pauli string \(c_j P_j\), prepare the ancilla in \(\ket{+}=(\ket{0}+\ket{1})/\sqrt2\), apply the interaction \(\exp(\ii\vartheta_j(\beta)P_j\otimes  Y)\), and measure it in the computational basis. Here \(Y\) is the ancilla Pauli operator and \(\vartheta_j\) is the interaction angle. {\color{black} We associate the computational-basis outcomes with the eigenvalues \(\pm1\), thereby defining the random variable 
\begin{equation}
s:\Omega\to\{\pm1\},\quad
s(\omega)= (-1)^\omega,
\quad
\Omega=\{0,1\}.  
\end{equation}
} 
For \(|\vartheta_j(\beta)|<\pi/4\), this operation is equivalent to
\begin{equation}
K_{\beta,j}(\omega)
=
\frac{1}{\sqrt{2}}
\left(
\cos\vartheta_j(\beta)\,\mathds{1}
+s(\omega)\sin\vartheta_j(\beta)\,P_j
\right),
\label{eq:exact-elementary-map}
\end{equation}
with 
\begin{equation}
\label{eq:exact-elementary-angle}
\vartheta_j(\beta)
=
-\arctan\!\left(\tanh\frac{\beta c_j}{2}\right).
\end{equation}

For every finite $\beta$, the conditional operator is proportional to the
imaginary-time exponential:
\begin{equation}
\label{eq:kraus_j}
K_{\beta,j}(\omega)
=
\sqrt{\frac{\cos 2\vartheta_j(\beta)}{2}}\,
\e^{-s(\omega)\frac{\beta}{2}c_j P_j}.
\end{equation}
Thus the two outcomes $\omega \in \Omega$ implement opposite imaginary-time directions with
the same scalar prefactor.

{\color{black}A sequence containing one interaction per term produces $m$ random
variables $s_1,\dots,s_m$ on the joint sample space
$\boldsymbol{\Omega}=\{0,1\}^m$:
\begin{equation}
    s_j:\boldsymbol{\Omega}\to\{\pm1\},
    \quad
 s_j(\boldsymbol{\omega})=(-1)^{\omega_j}.
\end{equation}
The corresponding random-signed Hamiltonian is
\begin{equation}
H({\boldsymbol{\omega}})=\sum_{j}s_j(\boldsymbol{\omega})c_j P_j
  \label{eq:signed-general-H}
\end{equation}
}
and the ordered conditional Kraus operator is
\begin{equation}
K_{\beta}(\boldsymbol\omega)
\propto
\prod_j\e^{-s_j(\boldsymbol\omega)\frac{\beta}{2} c_j P_j}.    \label{eq:one-step-map}
\end{equation}
Here, we have omitted the prefactor, which is independent of the measurement outcome and therefore cancels upon automatic renormalization.
Combining noncommuting terms gives a first-order product approximation to \(\e^{-\beta H(\boldsymbol{\omega})/2}\), which introduces the finite-step error assessed in Sec.~\ref{sec:results} and Appendix~\ref{app:checks}.

{\color{black}
The above derivation differs from Refs. \cite{Cuomo_2026,mao2023measurement} in that it does not impose an upper bound on the inverse temperature \(\beta\), which can therefore be chosen arbitrarily. This extension, however, comes with a statistical cost: as \(\beta\to\infty\), the probabilities of the measurement records become increasingly dependent on their sign configurations, potentially suppressing some records and making their sampling increasingly demanding. We discuss this effect in Appendix~\ref{app:prior-reuse}.
}

The family of signed Hamiltonians can be organized into a single operator on the tensor product of the spin Hilbert space and an auxiliary record register. Associate each bitstring \(\boldsymbol\omega\in\boldsymbol\Omega\) with the orthonormal record state \(\lvert\boldsymbol\omega\rangle\). The extended Hamiltonian is
\begin{equation}
\boldsymbol{\widetilde H}
=
\sum_{\boldsymbol\omega}
H({\boldsymbol\omega})
\otimes|\boldsymbol\omega\rangle\langle\boldsymbol{\omega}|.
\label{eq:extended-H}
\end{equation}
Each block contains the signed Hamiltonian associated with one record. The Gibbs state of this extended operator therefore has conditional spin states proportional to $e^{-\beta H(\boldsymbol{\omega})}$, with record weights proportional to the corresponding partition functions. These equilibrium weights generally differ from the
probabilities produced by the finite-depth circuit, which are
determined by the ordered Kraus operators in Eq.~\eqref{eq:one-step-map}. Appendix~\ref{app:extended-output} develops this distinction and gives the condition under which the two descriptions coincide.

To obtain thermal traces, the system input is the maximally mixed state \(\I/2^n\), as in Ref.~\cite{Cuomo_2026}. A purification prepares this state on the system by entangling it with a reference register that is subsequently ignored. Appendix~\ref{app:thermal-input} shows the preparation circuit and proves its equivalence to the uniformly weighted basis ensemble used in our numerical calculation. For the maximally mixed input, the probability of a complete record $\boldsymbol{\omega}$ and its normalized system state are
{\color{black}
\begin{equation}
\label{eq:record-state}
\operatorname{Pr}_{\beta} \{\boldsymbol{\omega}\} = \frac{\Tr K_{\beta}(\boldsymbol\omega) \mathds{1} K_{\beta}(\boldsymbol\omega)^{\dagger}}{2^n}
\end{equation}
}
and
\begin{equation}
\rho_{\beta}(\boldsymbol{\omega})=\frac{
K_\beta(\boldsymbol\omega)K_\beta(\boldsymbol\omega)^\dagger
}{
\operatorname{Tr}
K_\beta(\boldsymbol\omega)K_\beta(\boldsymbol\omega)^\dagger
}=\frac{e^{-{\beta}H({\boldsymbol{\omega}})}}{\Tr e^{-{\beta}H({\boldsymbol{\omega}})}} + \mathcal{O}(\beta^2).
\end{equation}



The above yields an approximation whose observables can be compared with that Gibbs target. Each run supplies the record \(\boldsymbol{\omega}\), which identifies the energy observable \(H(\boldsymbol{\omega})\), and a spin-measurement outcome used to estimate its expectation value in \(\rho_{\beta}(\boldsymbol{\omega})\):
{\color{black}
\begin{equation}
    \Tr H(\boldsymbol{\omega})\rho_\beta(\boldsymbol{\omega}) = \langle H(\boldsymbol{\omega})\rangle_\beta + \mathcal{O}(\beta^2).
\end{equation}
}

\subsection{Bond signs, frustration, and flux}

For the transverse-field Ising model, the record \(\boldsymbol{\omega}\) assigns a sign to each nearest-neighbor coupling and each transverse field. Let \(\mathcal V\) be the set of lattice sites and \(\mathcal E\) the set of bonds. The operators \(X_v\) and \(Z_v\) denote the Pauli operators at site \(v\). With positive coupling magnitudes \(J\) and \(h\), the signed Hamiltonian is
{\color{black}
\begin{equation}
\label{eq:signed-tfim}
H(\boldsymbol{\omega})=J\!\sum_{\{u,v\}\in \mathcal E}s_{\ell(u,v)}(\boldsymbol{\omega})Z_{u}Z_{v}+h\sum_{v\in \mathcal V}s_{v}(\boldsymbol{\omega})X_v.
\end{equation}
The index function $\ell$ maps every bond to a distinct record bit in $\{|\mathcal V|+1,\dots, |\mathcal V|+|\mathcal E|\}$.

A bond sign $s_{\ell(u,v)}(\boldsymbol{\omega})$  specifies whether the coupling favors antiparallel spins \((+1)\) or parallel spins \((-1)\) in the \(Z\) basis. Conjugation by \(Z_v\) reverses the sign of the transverse field at vertex \(v\) while leaving all bond couplings unchanged. Thus, the field signs can be changed independently by local unitary transformations without altering the bond-sign configuration.
Applying such a transformation maps any realization $\boldsymbol{\omega}$ to some  $\boldsymbol{\varpi}\in \boldsymbol{\Omega}$ satisfying
\begin{equation}
s_{v}(\boldsymbol{\varpi}) = 1, \quad s_{\ell(u,v)}(\boldsymbol{\omega}) =s_{\ell(u,v)}(\boldsymbol{\varpi})
\end{equation}
The same conjugation transforms each elementary Kraus factor and therefore the full ordered conditional operator. For the maximally mixed input, it leaves the record probability unchanged, so
\begin{equation} 
     \operatorname{Pr}_{\beta}\{\boldsymbol{\omega}\} = \operatorname{Pr}_{\beta}\{\boldsymbol{\varpi}\}.
\end{equation}

{\color{black}
Thus all field-sign strings are equally probable at fixed bond signs, even though their operators and states are generally different. This equiprobability property is useful when deriving the energy for a set of gauge-equivalent configurations.}

{\color{black}
After fixing the field signs, the local gauge transformation at site \(v\) acts on the signed Hamiltonian as
$X_vH(\boldsymbol{\omega})X_v$. This conjugation reverses the signs of all bonds incident on \(v\) and leaves the transverse-field terms unchanged. Configurations related by sequences of such local conjugations form a gauge class. The corresponding Hamiltonians are unitarily equivalent and therefore have the same spectrum, partition function, and thermal energy.
}

On a tree, one can choose the transformations successively from a root to make every bond positive, leaving a single class -- see Fig.~\ref{fig:workflow}(a). This recovers the sign-removal mechanism used for open chains in Ref.~\cite{Cuomo_2026}.

A closed loop retains information that this procedure cannot remove. For a cycle $\gamma \subseteq \mathcal E$, define its flux as the product of its bond signs
\begin{equation}
\Phi(\gamma,\boldsymbol{\omega}) = \prod_{\{u,v\}\in\gamma}s_{\ell(u,v)}(\boldsymbol\omega).
\label{eq:cycle-flux}
\end{equation}
On a square plaquette, \(\Phi=-1\) means that the preferred spin alignments cannot be satisfied on all four bonds simultaneously. The plaquette is frustrated -- see Fig.~\ref{fig:workflow}(b). Equal fluxes on an independent set of cycles are also sufficient for gauge equivalence~\cite{harary1953notion,mattis1976solvable,nishimori1981internal}. Appendix~\ref{app:gauge} gives the spanning-tree proof.

For a square lattice with open boundaries, the elementary plaquettes form a cycle basis $\Gamma$.
An \((a+1)\times (b+1)\) lattice has \(ab\) such plaquettes, each carrying a binary flux, and hence $2^{ab}$ gauge classes. 

Multiplying the signs, recorded in $\boldsymbol\omega$, around these plaquettes identifies a class
\begin{equation} [\boldsymbol\omega]_{\Gamma} = \{\boldsymbol\varpi\ |\ \Phi(\gamma,\boldsymbol\varpi) = \Phi(\gamma,\boldsymbol\omega),\ \forall \gamma \in \Gamma\}.
\end{equation}
The signed Hamiltonian \(H(\boldsymbol{\omega})\) defines its energy observable.

As in the measurement construction of Ref.~\cite{Cuomo_2026}, the bond and field terms are estimated in separate \(Z\)- and \(X\)-basis measurements, with the recorded signs included in classical processing. These signed energy contributions can then be pooled within each class. Appendix~\ref{app:gauge} gives the corresponding transformation rule. For other observables, the same procedure requires either gauge invariance or a transformation to a common representative frame, as described in Appendix~\ref{app:gauge}.


\section{Reconstructing quenched thermodynamics}
\label{sec:reconstruction}

Once records have been grouped by flux, their energies can be combined according to the desired distribution of couplings. Applying the same distribution at each temperature produces the energy curve of one quenched model, from which its free energy follows by integration.

\begin{figure*}[t]
    \centering
    \includegraphics[width=0.98\textwidth]{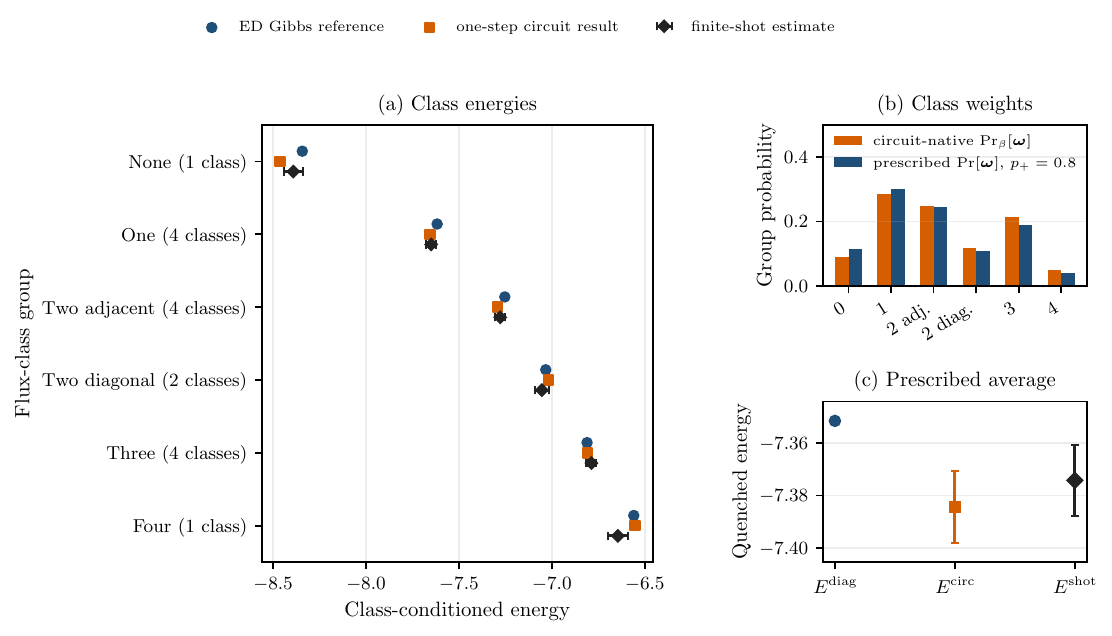}
    \caption{Reconstruction of the prescribed energy on the open \(3\times3\)
    lattice at \(\beta=0.75\).
    (a) Gibbs class energies from exact diagonalization (ED), exact conditional
    expectations of the one-step circuit, and finite-shot estimates, grouped by
    rotations and reflections of the square lattice. The labels specify the
    number and arrangement of frustrated plaquettes, for which \(\Phi=-1\).
    ``None'' means that all four plaquettes are unfrustrated, while ``Two
    adjacent'' and ``Two diagonal'' distinguish the two arrangements of two
    frustrated plaquettes (Fig.~\ref{fig:spatial-resolution}). Parentheses give
    the number of distinct flux classes in each symmetry group. For the uniform
    couplings and circuit ordering used here, symmetry-related classes have the
    same Gibbs energy and the same exact circuit energy, respectively. Their
    finite-shot data are pooled within each group.
    (b) Circuit probabilities \(\Pr_\beta[\boldsymbol\omega]\) and prescribed probabilities
    \(\Pr[\boldsymbol\omega]\) for independent positive bonds with probability \(p_+=0.8\),
    summed over the classes in each group.
    (c) Averaging with the prescribed prior gives \(E^{\mathrm{diag}}\),
    \(E^{\mathrm{circ}}\), and \(E^{\mathrm{shot}}\).
    Finite-shot error bars show one estimated standard error for the sampling
    prescription in Appendix~\ref{app:resources}. The error bar centered on
    \(E^{\mathrm{circ}}\) is the projected sampling uncertainty for that
    prescription, not an uncertainty in the exact circuit expectation.
    The displacement of \(E^{\mathrm{circ}}\) from \(E^{\mathrm{diag}}\)
    measures the finite-step bias.} 
    \label{fig:class-to-average}
\end{figure*}

\subsection{A prescribed disorder distribution}

Resolving the flux classes separates the measurement of their energies from the choice of disorder distribution. 
The circuit samples the class with probability
\begin{equation}
    \operatorname{Pr}_{\beta}[\boldsymbol{\omega}]
    =
    \sum_{\boldsymbol\varpi \in [\boldsymbol\omega]}\operatorname{Pr}_{\beta}\{\boldsymbol\varpi\},
    \label{eq:native-class-weight}
\end{equation}
which generally varies with temperature through the conditional-state norms in Eq.~\eqref{eq:record-state}. A quenched model instead specifies a distribution of fixed couplings. We denote its induced probability on the flux classes by \(\operatorname{Pr}[\boldsymbol{\omega}]\), and hold it fixed as the temperature changes.

The prescribed disorder average follows by weighting the resolved energies with this class prior,
\begin{equation}
    E(\beta)=\sum_{ [\boldsymbol{\omega}]} \operatorname{Pr}[\boldsymbol{\omega}]\langle H(\boldsymbol{\omega})\rangle_{\!\beta}.
    \label{eq:prescribed-energy}
\end{equation}

For Gibbs energies, this is the quenched mean energy. The thermal expectation is taken for fixed couplings before averaging over their distribution. Applying the same weights to circuit measurements estimates that quantity with the accuracy of the conditional imaginary-time evolution. {\color{black} Once the class means are available, choosing a distribution requires only classical postprocessing.} This also gives access to a family of disorder models from the same measurement data, without requiring a separate measurement campaign for each choice of disorder distribution.

The circuit induces a temperature-dependent distribution over gauge classes, \(\operatorname{Pr}_\beta[\boldsymbol{\omega}]\). Using the native circuit distribution to average the class-conditioned circuit energies gives

\begin{equation}
    \begin{aligned}
\tilde{E}^{\operatorname{circ}}(\beta)&=
\sum_{[\boldsymbol\omega]}
\operatorname{Pr}_\beta[\boldsymbol\omega]\,
\operatorname{Tr}
H(\boldsymbol\omega)\rho_\beta(\boldsymbol\omega)
\\
&=\sum_{[\boldsymbol{\omega}]} \operatorname{Pr}_\beta[\boldsymbol{\omega}]\langle H(\boldsymbol{\omega})\rangle_{\!\beta} + \mathcal{O}(\beta^2),
\end{aligned}
\end{equation}
which generally differs from Eq.~\eqref{eq:prescribed-energy}. To estimate the prescribed average, each class with \(\operatorname{Pr}[\boldsymbol{\omega}]>0\) must have nonzero sampling probability, and sufficient measurements must reach the classes carrying appreciable target weight. The overlap between \(\operatorname{Pr}[\boldsymbol{\omega}]\) and \(\operatorname{Pr}_{\beta}[\boldsymbol{\omega}]\) therefore controls the statistical cost, which directly impacts the measurement cost but does not change the target quantities being reconstructed, i.e.
\begin{equation}
    \label{eq:circ_energy}
    {E}^{\operatorname{circ}}(\beta)=\sum_{[\boldsymbol{\omega}]} \operatorname{Pr}[\boldsymbol{\omega}]\langle H(\boldsymbol{\omega})\rangle_{\!\beta} + \mathcal{O}(\beta^2),
\end{equation}
Appendix~\ref{app:estimators} gives both the estimator formed from separate class means and its importance-sampling counterpart.

Provided that every class with nonzero prescribed probability has nonzero circuit sampling probability, the circuit therefore supplies the prescribed disorder average of the class-conditioned energies.


\subsection{Free energy from the energy curve}

The energy as a function of temperature also determines the quenched free energy, providing a thermodynamic potential that distinguishes quenched from annealed disorder and can be reconstructed without directly measuring partition functions. {\color{black}
For any class $[\boldsymbol{\omega}]$, 
the Gibbs energy, valid for each element of the class, is given by
\begin{equation}
    \langle H(\boldsymbol{\omega})\rangle_\beta=-\frac{\partial}{\partial\beta}\log \Tr \e^{-\beta H(\boldsymbol{\omega})}.
    \label{eq:energy-logZ}
\end{equation}
The quenched free energy, obtained by averaging the Gibbs free energy over the fixed prior, is therefore
\begin{equation}
    {F}(\beta)=-\frac{1}{\beta}\sum_{ [\boldsymbol{\omega}]}\operatorname{Pr}[\boldsymbol{\omega}]\log\Tr\e^{-\beta H(\boldsymbol{\omega})},
    \label{eq:quenched-F}
\end{equation}

For comparison, annealed disorder allows the couplings to participate in the thermal average. For the same prior, its free energy is
\begin{equation}
    \tilde{F}(\beta) = -\frac{1}{\beta}
    \log\!\left(\sum_{ [\boldsymbol{\omega}]}\operatorname{Pr}[\boldsymbol{\omega}]\Tr\e^{-\beta H(\boldsymbol{\omega})}\right).
    \label{eq:annealed-F}
\end{equation}
}
The order of the logarithm and disorder average distinguishes the two thermodynamic quantities. In particular, the annealed class weights contain the temperature-dependent partition functions, whereas the quenched construction retains the prescribed \(\operatorname{Pr}[\boldsymbol{\omega}]\) throughout the energy integral.

The circuit supplies the averaged energy of Eq.~\eqref{eq:circ_energy}.
Substituting it into the thermodynamic integral gives
\begin{equation}
    F^{\rm circ}(\beta) = -\frac{n \log 2}{\beta} + \frac{1}{\beta}\int_{0}^{\beta}\!\dd\alpha\ E^{\rm circ}(\alpha).
\end{equation}
The normalization follows from \(\Tr\e^{-0 H(\boldsymbol{\omega})} = 2^n\).
The finite-depth bias is the integrated difference between the circuit and Gibbs energy curves, giving
\begin{equation}
    F^{\rm circ}(\beta) - F(\beta) = \frac{1}{\beta}\int_{0}^{\beta}\!\dd\alpha\,\bigl( E^{\rm circ}(\alpha) - E(\alpha)\bigr).
\end{equation}
Sampling fluctuations and numerical quadrature introduce additional errors when the integral is estimated from data. Keeping these contributions separate allows the numerical study to assess how faithfully the circuit reconstructs the target free energy.

Replacing the circuit expectations by finite-shot estimates yields  $E^{\rm shot}(\beta)$ and $F^{\rm shot}(\beta)$.
}

\begin{figure*}[t]
    \centering
    \includegraphics[width=0.94\textwidth]{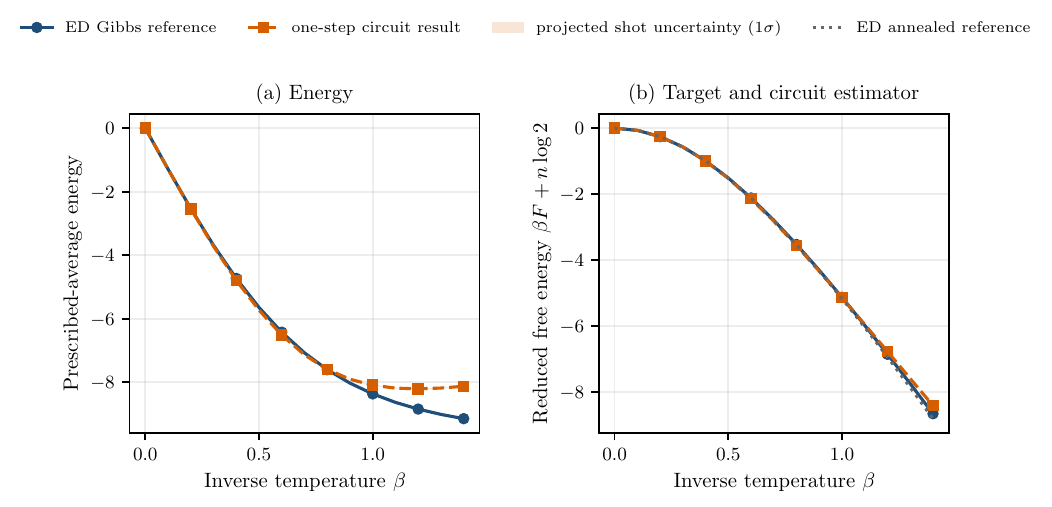}
    \caption{Quenched free energy from thermodynamic integration for the fixed
    prior \(p_+=0.8\).
    (a) Gibbs energies \(E^{\mathrm{diag}}(\beta)\) and exact one-step circuit
    energies \(E^{\mathrm{circ}}(\beta)\), averaged with the same prescribed
    class weights \(\Pr[\boldsymbol\omega]\).
    (b) The corresponding reduced free energies
    \(\beta F^{\mathrm{diag}}+n\log2\) and
    \(\beta F^{\mathrm{circ}}+n\log2\), together with the ED annealed reference
    \(\beta\widetilde F+n\log2\).
    The circuit curve is obtained by Simpson integration on fifteen
    inverse-temperature points, with markers displayed at every second point.
    The bands show the projected one-standard-error sampling uncertainty for
    the prescription in Appendix~\ref{app:resources}, propagated through the
    integral in panel (b). The central circuit curves are calculated from exact
    conditional expectations. No finite-shot free-energy curve is plotted.
    Their separation from the Gibbs references measures finite-step bias,
    with the smaller quadrature contribution assessed in Appendix~\ref{app:checks}.}
    \label{fig:quenched-free-energy}
\end{figure*}

\Needspace{10\baselineskip}
\section{Numerical reconstruction on an open square lattice}
\label{sec:results}

{\color{black}
For the experiments, we prescribe independent Bernoulli distributions with parameter \(\rm p_\texttt{+}\), such that each bond is positive with probability \(\rm p_\texttt{+}\) and negative with probability \(1-\rm p_\texttt{+}\).
The probabilities are determined by
\begin{equation}
    \label{eq:prescribed-record-prior}
\operatorname{Pr}\{\boldsymbol\omega\}
= 
\frac{1}{2^n}\;\rm{ p}_\texttt{+}\phantom{)\!}^{\!\sum_{\{u,v\}}\!\neg \omega_{\ell(u,v)}}
(1-\rm{p}_\texttt{+})^{\sum_{\{u,v\}} \omega_{\ell(u,v)}},
\end{equation}
with $\neg$ being the Boolean complement.
Summing over the field bits recovers the prescribed independent-bond distribution. Its class weights are 
\begin{equation}
    \operatorname{Pr}[\boldsymbol\omega] = \sum_{\boldsymbol\varpi \in [\boldsymbol{\omega}]} \operatorname{Pr}\{\boldsymbol\varpi\}.
\end{equation}
These weights are applied during classical postprocessing and the circuit continues to sample records according to $\operatorname{Pr}_{\beta}$.
}

The open \(3\times3\) lattice provides a small system in which both the amount and the arrangement of frustration can vary. Its nine spins are joined by twelve bonds, forming four independent plaquettes and sixteen flux classes. Under the spatial symmetries of the square, these classes fall into six groups. In particular, two frustrated plaquettes can be adjacent or diagonal. We use \(J=h=\pi/4\) and the independent-bond distribution with \(\rm p_{\texttt{+}}=0.8\). All energies and inverse temperatures below use the same coupling units.

The finite-shot calculation uses an ideal matrix-product-state circuit simulator and one product-formula step with the exact elementary angles in Eq.~\eqref{eq:exact-elementary-angle}. The circuit applies the $ZZ$ interactions before the $X$ interactions and uses nine system qubits and twenty-one ancillas, which record the signs of the twelve bonds and nine transverse fields. The maximally mixed input is implemented as a uniformly weighted computational-basis ensemble. This gives the same reduced system state as purification, while omitting the nine reference qubits and their preparation gates from the numerical circuits (Appendix~\ref{app:thermal-input}). Measurements in the \(Z\) and \(X\) bases determine the signed bond and field contributions to the energy. The sampling prescription is given in Appendix~\ref{app:resources}. Separately, we calculate exact conditional circuit expectations from the ordered Kraus operators and Gibbs energies obtained by ED.

At \(\beta=0.75\), the finite-shot calculation resolves the energy differences between the six symmetry groups -- see Fig.~\ref{fig:class-to-average}(a). Within each group, the ED and exact circuit energies agree across symmetry-related classes to numerical precision. They agree to within \(4.1\times10^{-14}\) for ED and \(1.8\times10^{-15}\) for the exact circuit calculation, permitting their data to be pooled. The resolved energies also distinguish the arrangement of frustration. Configurations with two frustrated plaquettes have higher energy when the plaquettes are diagonal rather than adjacent, with differences of $0.220213$ for the Gibbs target and $0.275646$ for the one-step circuit. This shows that resolving the full flux class retains information about the spatial organization of frustration that would be lost if configurations were characterized only by the number of frustrated plaquettes.

Weighting the resulting class energies with the prescribed prior gives the target
\begin{equation}
    E^{\rm diag}(\beta)=-7.351629
\end{equation}
and the exact one-step circuit expectation 
\begin{equation}
    E^{\rm circ}(\beta)=-7.384441.
\end{equation}
The finite-shot estimate is 
\begin{equation}
    E^{\rm shot}(\beta) = -7.374307\pm0.013585,
\end{equation}
where the uncertainty is one estimated standard error. This estimate agrees with the exact circuit expectation within its sampling uncertainty, whose projected standard error is \(0.013581\). The comparison of exact expectations isolates the finite-step displacement from the Gibbs target, while the finite-shot result confirms the reconstruction of the circuit's prescribed average within the stated uncertainty.

The difference between the sampling and disorder distributions has a visible effect on the reconstructed energy. Figure~\ref{fig:class-to-average}(b) compares their weights, and panel (c) shows the prescribed average obtained from the resolved energies. Holding the ED class energies fixed but weighting them by \(\operatorname{Pr}_\beta[\boldsymbol{\omega}]\) gives \(-7.302453\), compared with \(-7.351629\) under \(\operatorname{Pr}[\boldsymbol{\omega}]\). The change comes entirely from the weighting. This difference demonstrates why retaining the class-resolved data is essential. Using the circuit’s native frequencies would answer a different thermodynamic question from the one defined by the prescribed disorder model.

Once resolved, the data can be reused to vary the disorder model. Appendix~\ref{app:estimators} applies the same class energies to a range of \(\rm p_{\texttt{+}}\) values and quantifies the associated sampling cost.

Repeating the energy calculation at fifteen equally spaced inverse temperatures from \(\beta=0\) to \(1.4\) gives the curves in Fig.~\ref{fig:quenched-free-energy}(a). Simpson integration of the exact circuit energies then produces the free-energy estimate in panel (b). At $\beta = 1.4$, the exact target and the integrated circuit estimate are
\begin{align}
  F^{\rm diag}(1.4)&=-10.636534,\\
    F^{\rm circ}(1.4)&=-10.459832\pm0.003912.
\end{align}
The uncertainty shown here is the projected sampling error of the estimator propagated through the integral. The central value was calculated from exact circuit expectations and the finite-shot circuit calculation is performed at \(\beta=0.75\), rather than at every point of the temperature sweep. The bias of \(+0.176702\) is substantially larger and reflects the accumulated difference between the one-step circuit and Gibbs energy curves. The ED quenched free energy also exceeds the annealed value by \(0.107066\), showing the thermodynamic consequence of keeping the disorder distribution fixed during thermal averaging. The reconstruction procedure is statistically well controlled in this calculation, while the present one-step approximation limits its quantitative accuracy.

Numerical integration contributes little to this discrepancy. Integrating the ED energy curve on the fifteen-point grid changes the exact endpoint by \(-2.20\times10^{-5}\), and refining the circuit grid to twenty-nine points changes its estimate by \(-1.84\times10^{-5}\). Both corrections are much smaller than the finite-step bias. Separate checks on \(2\times2\) and \(2\times3\) lattices verify the elementary conditional map and gauge covariance, while twelve independent sampling repeats test the uncertainty estimate. Appendix~\ref{app:checks}
presents these checks and the additional spatial information contained in
the resolved classes. 

Together, these results demonstrate that the class-resolved measurement record is sufficient to reconstruct a prescribed quenched energy and free energy, while also identifying the finite-step approximation—not the disorder reweighting, sampling, or numerical integration—as the dominant limitation of the present implementation.

\FloatBarrier
\section{Discussion and Outlook}
\label{sec:discussion}

Ancilla records provide a way to reconstruct disorder-averaged thermodynamics from measurements resolved by gauge class. The key advantage is that the measurement record retains enough information to postpone the choice of disorder distribution until classical postprocessing, separating the sampling strategy of the circuit from the physical disorder model. This also gives access to energies that are typically challenging to calculate, such as those of quenched systems \cite{Bravyi2022QuantumHamiltonianComplexity}. 

The local sign transformations that identify equivalent records also determine which information survives. A tree has one class, while cycles retain the fluxes associated with frustration. Measuring the energy within each class makes it possible to choose the disorder distribution when combining the data. The \(3\times3\) calculation demonstrates this procedure for a prescribed quenched energy and extends it to a free-energy estimate through thermodynamic integration.

The resolved classes also retain information about the spatial organization of frustration. The energy difference between adjacent and diagonal arrangements shows that the number of frustrated plaquettes alone does not determine the class energy. Conditioning on the full flux pattern preserves these distinctions when the class energies are combined with a prescribed disorder weights. 

The present simulations identify the accuracy and sampling requirements of this approach. For the chosen one-step circuit, the free-energy error is dominated by the product approximation, while statistical uncertainty and quadrature error are smaller. Reducing that bias motivates deeper or higher-order constructions in which each retained record still identifies a fixed signed Hamiltonian. A trajectory with signs that change between steps describes a time-dependent sequence, so extending the static-disorder reconstruction requires control of the record history as well as the product-formula error. Appendix~\ref{app:map} makes this condition explicit. Extending the reconstruction to larger $\beta$, and hence lower temperatures, will require control of this approximation together with sufficient sampling of the classes important to the chosen disorder model. 

For larger lattices, the main statistical question is how well the circuit samples the classes important to the chosen disorder model. The number of classes grows exponentially with the number of independent cycles, although identifying an individual record on an open square lattice requires only products around local plaquettes. The  nominal effective sample fraction, defined in Eq.~\eqref{eq:app-neff}, quantifies the overlap between the prescribed class weights and the circuit sampling distribution. For the demonstrated prior, \(N_{\rm eff}/N_{\rm shots}\approx0.99\) at \(\beta=0.75\), decreasing to approximately \(0.75\) at \(\beta=1.4\). 
This decrease occurs at fixed prescribed disorder and reflects worse overlap with the circuit distribution at the larger inverse temperature. Symmetry pooling, priors with suitable sampling overlap, and direct importance-weighted estimates offer ways to study selected averages without estimating every class mean separately.

The reconstruction developed here separates the circuit's sampling distribution from the disorder distribution defining the thermodynamic model. Resolving ancilla records by gauge class allows the measured energies to be combined with prescribed disorder weights and reused for different coupling distributions. The open-square calculation demonstrates this procedure for quenched energy and extends it to a free-energy estimate through thermodynamic integration, while quantifying the finite-step bias. The retained flux information also distinguishes spatial arrangements of frustration, connecting the thermodynamic reconstruction to physical information carried by the measurement record, and preserving such information beyond the disorder average. Therefore, the same measurement record may also support more detailed, class-resolved analysis of frustration. 

Extending this reconstruction to larger frustrated spin systems will require circuits that improve conditional thermal accuracy while retaining interpretable records, together with sampling strategies suited to the prescribed disorder distribution.

\section{Acknowledgments}

The collaboration between Cuomo and Adisa began through the
Qiskit Advocate Mentorship Program (QAMP 2025)
\cite{CuomoQAMP2025}. We thank the program organizers for bringing
the authors together and supporting their collaboration.
All authors contributed voluntarily as Qiskit Advocates, and
the work received no external funding. 

\section*{Data and code availability}

The simulation code, files and scripts used to regenerate the numerical
results presented in this manuscript are publicly available in
the accompanying GitHub repository \cite{AdisaCuomoGaugeITE}.

\bibliographystyle{unsrt}
\bibliography{refs}

\begin{thebibliography}{10}

\bibitem{Motta2020QITE}
Mario Motta, Chong Sun, Andrew T.~K. Tan, Matthew~J. O'Rourke, Erika Ye, Austin~J. Minnich, Fernando G. S.~L. Brandão, and Garnet Kin-Lic Chan.
\newblock Determining eigenstates and thermal states on a quantum computer using quantum imaginary time evolution.
\newblock {\em Nature Physics}, 16:205--210, 2020.

\bibitem{McArdle2019VITE}
Sam McArdle, Tyler Jones, Suguru Endo, Ying Li, Simon~C. Benjamin, and Xiao Yuan.
\newblock Variational ansatz-based quantum simulation of imaginary time evolution.
\newblock {\em npj Quantum Information}, 5:75, 2019.

\bibitem{Yuan2019VQS}
Xiao Yuan, Suguru Endo, Qi~Zhao, Ying Li, and Simon~C. Benjamin.
\newblock Theory of variational quantum simulation.
\newblock {\em Quantum}, 3:191, 2019.

\bibitem{mao2023measurement}
Yuping Mao, Manish Chaudhary, Manikandan Kondappan, Junheng Shi, Ebubechukwu~O Ilo-Okeke, Valentin Ivannikov, and Tim Byrnes.
\newblock Measurement-based deterministic imaginary time evolution.
\newblock {\em Physical Review Letters}, 131(11), 2023.

\bibitem{kondappan2023imaginary}
Manikandan Kondappan, Manish Chaudhary, Ebubechukwu~O Ilo-Okeke, Valentin Ivannikov, and Tim Byrnes.
\newblock Imaginary-time evolution with quantum nondemolition measurements: Multiqubit interactions via measurement nonlinearities.
\newblock {\em Physical Review A}, 107(4), 2023.

\bibitem{Chen2023Thermal}
Chi-Fang Chen, Michael~J. Kastoryano, Fernando G. S.~L. Brandão, and András Gilyén.
\newblock Quantum thermal state preparation.
\newblock {\em arXiv preprint arXiv:2303.18224}, 2023.

\bibitem{Cuomo_2026}
Daniele Cuomo and Robert van Leeuwen.
\newblock Quantum simulation in imaginary time for gauge-invariant models.
\newblock {\em Physica Scripta}, 101(13):135106, mar 2026.

\bibitem{wegner1971duality}
Franz~J Wegner.
\newblock Duality in generalized ising models and phase transitions without local order parameters.
\newblock {\em Journal of Mathematical Physics}, 12(10):2259--2272, 1971.

\bibitem{toulouse1987theory}
G~Toulouse et~al.
\newblock Theory of the frustration effect in spin glasses: I.
\newblock {\em Spin Glass Theory and Beyond: An Introduction to the Replica Method and Its Applications}, 9:99, 1987.

\bibitem{morita2005gauge}
Satoshi Morita, Yukiyasu Ozeki, and Hidetoshi Nishimori.
\newblock Gauge theory for quantum spin glasses.
\newblock {\em Journal of the Physical Society of Japan}, 75(1), 2006.

\bibitem{Bravyi2022QuantumHamiltonianComplexity}
Sergey Bravyi, Anirban Chowdhury, David Gosset, and Pawel Wocjan.
\newblock Quantum hamiltonian complexity in thermal equilibrium.
\newblock {\em Nature Physics}, 18:1367--1370, 2022.

\bibitem{sinclair_et_al:LIPIcs.TQC.2026.1}
Alistair Sinclair and Thuy-Duong Vuong.
\newblock {A Sharp Computational Phase Transition for the Partition Function of the Transverse-Field Ising Model}.
\newblock In {\em 21st Conference on the Theory of Quantum Computation, Communication and Cryptography (TQC 2026)}, volume 389 of {\em Leibniz International Proceedings in Informatics (LIPIcs)}, pages 1:1--1:20. Schloss Dagstuhl -- Leibniz-Zentrum fuer Informatik, 2026.

\bibitem{harary1953notion}
Frank Harary.
\newblock On the notion of balance of a signed graph.
\newblock {\em Michigan Mathematical Journal}, 2(2):143--146, 1953.

\bibitem{mattis1976solvable}
DC~Mattis.
\newblock Solvable spin systems with random interactions.
\newblock {\em Physics Letters A}, 56(5):421--422, 1976.

\bibitem{nishimori1981internal}
Hidetoshi Nishimori.
\newblock Internal energy, specific heat and correlation function of the bond-random ising model.
\newblock {\em Progress of Theoretical Physics}, 66(4):1169--1181, 1981.

\bibitem{CuomoQAMP2025}
Daniele Cuomo.
\newblock Quantum simulation in imaginary time for gauge-invariant models.
\newblock \url{https://github.com/qiskit-advocate/qamp-2025/issues/5}, 2025.
\newblock Qiskit Advocate Mentorship Program (QAMP 2025), project issue no.~5.

\bibitem{AdisaCuomoGaugeITE}
Ibukunoluwa Adisa and Daniele Cuomo.
\newblock {Gauge ITE}: Gauge-class-resolved imaginary-time simulation.
\newblock \url{https://github.com/danielecuomo/GaugeModelSimulation}, 2026.
\newblock GitHub repository.

\bibitem{trotter1959product}
Hale~F Trotter.
\newblock On the product of semi-groups of operators.
\newblock {\em Proceedings of the American Mathematical Society}, 10(4):545--551, 1959.

\bibitem{suzuki1976generalized}
Masuo Suzuki.
\newblock Generalized trotter's formula and systematic approximants of exponential operators and inner derivations with applications to many-body problems.
\newblock {\em Communications in Mathematical Physics}, 51(2):183--190, 1976.

\bibitem{childs2021theory}
Andrew~M Childs, Yuan Su, Minh~C Tran, Nathan Wiebe, and Shuchen Zhu.
\newblock Theory of trotter error with commutator scaling.
\newblock {\em Physical Review X}, 11(1):011020, 2021.

\end{thebibliography}

\clearpage
\onecolumngrid
\appendix

\section{Conditional maps and thermal-state preparation}
\label{app:map}

The conditional map follows directly from the ancilla interaction. This appendix derives its exact imaginary-time parameter, describes the input needed for thermal traces, and relates the measured circuit to the extended Hamiltonian. The final subsection treats the record history in repeated-step constructions.

\subsection{Exact elementary map}

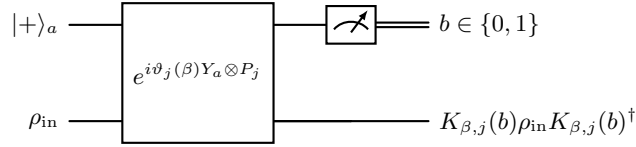
\begin{figure}[htbp]
\centering
\begin{quantikz}[row sep=0.7cm,column sep=0.7cm]
\lstick{\(|+\rangle_a\)} & \gate[2]{e^{i\vartheta_j(\beta)Y_a\otimes P_j}} & \meter{} & \setwiretype{c}\rstick{\(b\in\{0,1\}\)} \\
\lstick{\(\rho_{\mathrm{in}}\)} & & \qw & \rstick{\(K_{\beta,j}(b)\rho_{\mathrm{in}}K_{\beta,j}(b)^\dagger\)}
\end{quantikz}
\caption{Elementary measured interaction in the ancilla-resolved
construction of Ref.~\cite{Cuomo_2026}, with the Kraus operator
\(K_{\beta,j}(b)\) defined in the main text. The measured ancilla bit
\(b\in\{0,1\}\) sets \(s(b)=(-1)^b\). Both outcomes are retained.
Here \(\rho_{\mathrm{in}}\) is the system state before this interaction.
The displayed output \(K_{\beta,j}(b)\rho_{\mathrm{in}}
K_{\beta,j}(b)^\dagger\) is unnormalized, and its trace is the outcome
probability. The ancilla is drawn above the system, so the interaction is
written with the ancilla tensor factor first.}
\label{fig:elementary-circuit}
\end{figure}

The interaction in Fig.~\ref{fig:elementary-circuit} uses the outcome
convention \(s(b)=(-1)^b\) for a single measured bit \(b\).
Its elementary conditional operator is \(K_{\beta,j}(b)\), as defined in
the main text. In the compiled circuit, a \(Z_a\otimes P_j\) interaction
is followed by an ancilla rotation \(R_x(\pi/2)\). Since this rotation
sends \(Z_a\) to \(-Y_a\) and preserves \(|+\rangle_a\) up to a phase,
it implements the same conditional maps as the displayed interaction.
Thus the analytical labels coincide with the measured-bit convention used
in the numerical calculation. In a complete record, the bit for this term
is \(b=\omega_j\), so \(s(b)=s_j(\boldsymbol\omega)\).

The algebraic identity \(P_j^2=\mathbf{1}\) reduces the Pauli exponential
to two terms,
\begin{equation}
 e^{s(b)\lambda P_j}
 =\cosh\lambda\,\mathbf{1}+s(b)\sinh\lambda\,P_j.
 \label{eq:app-pauli-exp}
\end{equation}
Matching this expression with the elementary Kraus operator gives
\begin{equation}
 \tanh\lambda=\tan\vartheta_j(\beta),\qquad
 \lambda=\operatorname{arctanh}\left(\tan\vartheta_j(\beta)\right)
 =-\frac{\beta c_j}{2}.
 \label{eq:app-eta}
\end{equation}
For \(|\vartheta_j(\beta)|<\pi/4\), matching the identity coefficient gives
the outcome-independent prefactor
\begin{equation}
 \kappa_j(\beta)=\sqrt{\frac{\cos(2\vartheta_j(\beta))}{2}}.
\end{equation}
Consequently, the angle
\begin{equation}
 \vartheta_j(\beta)=-\arctan\!\left[\tanh\!\left(\frac{\beta c_j}{2}\right)\right]
 \label{eq:app-angle}
\end{equation}
gives \(K_{\beta,j}(b)=\kappa_j(\beta)e^{-s(b)\beta c_jP_j/2}\),
establishing the exact elementary exponential in the main text.
Here \(\lambda\) is only the scalar parameter used in this derivation.

\subsection{One layer and the thermal input}
\label{app:thermal-input}

Fix an ordering of the \(m\) Hamiltonian terms and use one ancilla
interaction per term. Define
\(\kappa(\beta)=\prod_{j=1}^{m}\kappa_j(\beta)\). The complete
conditional operator is
\begin{equation}
 \begin{aligned}
 K_\beta(\boldsymbol\omega)
 &=\prod_j^{\leftarrow}K_{\beta,j}(\omega_j)\\
 &=\kappa(\beta)\prod_j^{\leftarrow}
 e^{-\beta s_j(\boldsymbol\omega)c_jP_j/2}.
 \end{aligned}
 \label{eq:app-one-layer}
\end{equation}
The arrow orders factors so that interactions applied earlier act on the
right. It does not prescribe the chronological order from the numerical
record-bit labels. The first equality is the exact conditional circuit operation, including
its physical normalization. Dividing by \(\kappa(\beta)\) gives a
first-order Lie product approximation to \(e^{-\beta H(\boldsymbol\omega)/2}\),
which is exact when all terms commute. The prefactor cancels from the
normalized conditional state but must be retained when calculating
\(\Pr_\beta\{\omega\}=2^{-n}\operatorname{Tr}
K_\beta(\boldsymbol\omega)K_\beta(\boldsymbol\omega)^\dagger\).

The maximally mixed input is required because thermal observables are
traces. Following the maximally entangled input construction of
Ref.~\cite{Cuomo_2026}, introduce an \(n\)-qubit reference register \(R\)
alongside the \(n\)-qubit system \(S\). A Hadamard on each system qubit
followed by a CNOT from that qubit to its reference partner prepares
\(n\) Bell pairs -- see Fig.~\ref{fig:purified-protocol}(a). Their joint state is
\begin{equation}
 |\Phi\rangle=\frac{1}{\sqrt{2^n}}
 \sum_{q=0}^{2^n-1}|q\rangle_S|q\rangle_R,\qquad
 \operatorname{Tr}_R|\Phi\rangle\langle\Phi|
 =\frac{\mathbf{1}}{2^n}.
 \label{eq:app-purification}
\end{equation}
The complete computational basis provides an equivalent average,
\begin{equation}
 \frac{1}{2^n}\sum_{q=0}^{2^n-1}|q\rangle\langle q|
 =\frac{\mathbf{1}}{2^n}.
 \label{eq:app-basis-average}
\end{equation}
The equivalence persists through the conditional circuit. Let
\(K_\beta(\boldsymbol\omega)\) act on \(S\), with the reference register left
untouched. The unnormalized reduced state associated with record
\(\omega\) is
\begin{equation}
 \begin{aligned}
 &\operatorname{Tr}_R\!
 (K_\beta(\boldsymbol{\omega})\otimes\I_R)|\Phi\rangle\langle\Phi|
 (K_\beta(\boldsymbol{\omega})^\dagger\otimes\I_R)\\
 &\qquad=\frac{K_\beta(\boldsymbol{\omega})K_\beta(\boldsymbol{\omega})^\dagger}{2^n}
 =\frac{1}{2^n}\sum_{q=0}^{2^n-1}
 K_\beta(\boldsymbol{\omega})|q\rangle\langle q|K_\beta(\boldsymbol{\omega})^\dagger.
 \end{aligned}
 \label{eq:app-input-equivalence}
\end{equation}
Taking its trace gives the same record probability for both preparations. Normalizing it gives the same conditional system state. Consequently, all joint probabilities of the record and subsequent system measurements agree when the reference is ignored, including at finite product depth. In the basis implementation, joint records are pooled before conditioning. Equally averaging the separately normalized conditional states would generally give different weights.

\begin{figure}[htbp]
\centering
\textbf{(a)} Bell-pair preparation, repeated for \(j=1,\ldots,n\)\par\medskip
\begin{quantikz}[row sep=0.45cm,column sep=0.65cm]
\lstick{\(S_j\quad|0\rangle\)} & \gate{H} & \ctrl{1} & \qw \\
\lstick{\(R_j\quad|0\rangle\)} & \qw & \targ{} & \qw
\end{quantikz}
\par\bigskip
\textbf{(b)} Ancilla-resolved evolution with the purified input\par\medskip
\begin{quantikz}[row sep=0.55cm,column sep=0.6cm]
\lstick{\(A\quad|+\rangle^{\otimes m}\)} & \gate[2]{U_{AS}} & \meter{} & \setwiretype{c}\rstick{\(\boldsymbol{\omega}\in\{0,1\}^m\)} \\
\lstick{\(S\)} & & \meter{B} & \setwiretype{c}\rstick{spin outcomes} \\
\lstick{\(R\)} & \qw & \qw & \rstick{ignored}
\end{quantikz}
\caption{Purification and measurement of the thermal input.
(a) Each Hadamard--CNOT pair prepares \((|00\rangle+|11\rangle)/\sqrt2\)
on \(S_jR_j\). The \(n\) pairs form \(|\Phi\rangle_{SR}\) in
Eq.~\eqref{eq:app-purification}.
(b) Each wire represents a register. \(A\) contains \(m\) record qubits,
while \(S\) and \(R\) contain \(n\) qubits each. The registers \(S,R\)
enter jointly in the state prepared in (a). The unitary \(U_{AS}\) is the
ordered sequence of elementary interactions in Fig.~\ref{fig:elementary-circuit}.
Computational-basis measurements of \(A\) give the complete bit record
\(\boldsymbol{\omega}\), from which the signs \(s_j(\boldsymbol{\omega})=(-1)^{\omega_j}\) are
obtained. Measurements of \(S\) in basis \(B=Z\) or \(X\) supply the
signed bond or field energy contribution. The reference register is neither
acted on nor measured after preparation. Replacing the purification by the
uniformly weighted basis ensemble on \(S\) gives identical ideal joint
record--system probabilities by Eq.~\eqref{eq:app-input-equivalence}.}
\label{fig:purified-protocol}
\end{figure}

The \(3\times3\) calculation uses Eq.~\eqref{eq:app-basis-average} to omit the nine reference qubits and their entangling preparation gates. This reduces circuit width but requires \(2^n\) separate input settings when the full basis is enumerated, so it is a small-system numerical choice rather than an efficient preparation prescription for arbitrary \(n\). Purification, or uniform random selection of basis states, avoids exhaustive enumeration. The exact density-operator identity does not imply identical finite-sample fluctuations. The numerical calculation fixes equal shot counts for every input, and its uncertainty estimate accounts for that sampling design. A single fixed pure input instead changes the density operator and generally biases the thermal trace, as the comparison in Appendix~\ref{app:checks} illustrates.

\subsection{The extended Hamiltonian and the measured output}
\label{app:extended-output}

The extended Hamiltonian \(\boldsymbol{\widetilde H}\) defined in the main text places
every signed spin Hamiltonian in a separate record block.
Exponentiation preserves this block structure, so its partition function
and equilibrium record weights are
\begin{equation}
 \begin{aligned}
 \operatorname{Tr}e^{-\beta\boldsymbol{\widetilde H}}
 &=\sum_{\boldsymbol{\omega}\in\boldsymbol{\Omega}}\operatorname{Tr}e^{-\beta H(\boldsymbol{\omega})},\\
 w_\beta^{\mathrm{ext}}(\boldsymbol{\omega})
 &=\frac{\operatorname{Tr}e^{-\beta H(\boldsymbol{\omega})}}
 {\sum_{\boldsymbol{\varpi}\in\boldsymbol{\Omega}}\operatorname{Tr}e^{-\beta H(\boldsymbol{\varpi})}}.
 \end{aligned}
 \label{eq:app-formal-weights}
\end{equation}
These weights describe the Gibbs state of the formal extended operator.
They also describe the measured record distribution if every conditional
map has the form
\begin{equation}
 K_\beta(\boldsymbol{\omega})=a_\beta e^{-\beta H(\boldsymbol{\omega})/2}V_{\boldsymbol{\omega}},
 \qquad V_{\boldsymbol{\omega}} V_{\boldsymbol{\omega}}^\dagger=\I,
 \label{eq:app-gibbs-condition}
\end{equation}
where \(a_\beta\) is independent of the record and \(V_{\boldsymbol{\omega}}\) is
unitary. Record-dependent scalar phases can be absorbed into
\(V_{\boldsymbol{\omega}}\). Under this condition,
\(K_\beta(\boldsymbol{\omega})K_\beta(\boldsymbol{\omega})^\dagger
=|a_\beta|^2e^{-\beta H(\boldsymbol{\omega})}\), giving both the Gibbs conditional
state and its partition-function record weight. Finite product depth
generally changes this operator and its trace. The circuit distribution
must therefore be calculated from the physically normalized Kraus
operators as \(\Pr_\beta\{\boldsymbol{\omega}\}=2^{-n}\operatorname{Tr}
K_\beta(\boldsymbol{\omega})K_\beta(\boldsymbol{\omega})^\dagger\), rather than identified
with \(w_\beta^{\mathrm{ext}}(\boldsymbol{\omega})\).

An exact extended description of the circuit itself is available by
defining, for \(\beta>0\),
\begin{equation}
 \begin{aligned}
 H^{\mathrm{op}}(\boldsymbol{\omega},\beta)
 &=-\frac{1}{\beta}\log\bigl(K_\beta(\boldsymbol{\omega})K_\beta(\boldsymbol{\omega})^\dagger\bigr),\\
 \boldsymbol{\widetilde H}^{\mathrm{op}}(\beta)
 &=\sum_{\boldsymbol{\omega}\in\boldsymbol{\Omega}}H^{\mathrm{op}}(\boldsymbol{\omega},\beta)
 \otimes|\boldsymbol{\omega}\rangle\langle\boldsymbol{\omega}|.
 \end{aligned}
 \label{eq:app-operational-H}
\end{equation}
The finite-angle maps considered here are invertible, so the operator
logarithm is well defined. The joint state of the system and the measured
record is then
\begin{equation}
 \sum_{\boldsymbol{\omega}\in\boldsymbol{\Omega}}\Pr_\beta\{\boldsymbol{\omega}\}\rho_\beta(\boldsymbol{\omega})
 \otimes|\boldsymbol{\omega}\rangle\langle\boldsymbol{\omega}|
 =\frac{e^{-\beta\boldsymbol{\widetilde H}^{\mathrm{op}}(\beta)}}
 {\operatorname{Tr}e^{-\beta\boldsymbol{\widetilde H}^{\mathrm{op}}(\beta)}}.
 \label{eq:app-operational-state}
\end{equation}
For the physically normalized Kraus operators, the denominator in
Eq.~\eqref{eq:app-operational-state} equals \(2^n\), by the
Kraus completeness relation
\[
 \sum_{\boldsymbol{\omega}\in\boldsymbol{\Omega}}K_\beta(\boldsymbol{\omega})^\dagger K_\beta(\boldsymbol{\omega})
 =\I.
\]
This representation incorporates the actual conditional maps, including
their finite-step error. Its blocks agree with \(H(\boldsymbol{\omega})\) up to a
common scalar shift when Eq.~\eqref{eq:app-gibbs-condition} holds.
Since \(\boldsymbol{\widetilde H}^{\mathrm{op}}\) generally depends on \(\beta\),
this representation alone does not supply the fixed-Hamiltonian
thermodynamic identity used for the target free energy.

The formal operator \(\boldsymbol{\widetilde H}\) also has a simple auxiliary
symmetry. Complementing every bit, \(\boldsymbol{\omega}\mapsto\overline{\boldsymbol{\omega}}\),
reverses every sign, so \(H(\overline{\boldsymbol{\omega}})=-H(\boldsymbol{\omega})\).
The unitary \(|\boldsymbol{\omega}\rangle\mapsto|\overline{\boldsymbol{\omega}}\rangle\) therefore
sends \(\boldsymbol{\widetilde H}\) to \(-\boldsymbol{\widetilde H}\), making its spectrum
symmetric about zero. This transformation acts on the record register.
Gauge equivalence of physical spin Hamiltonians is instead determined by
the cycle fluxes in Appendix~\ref{app:gauge}.

\subsection{Repeated steps and record histories}
\label{app:record-histories}

For \(r\) steps of size \(\delta=\beta/r\), retain the same
complete record \(\boldsymbol{\omega}\) at every step. The resulting
conditional operator is \(K_{\beta,r}(\boldsymbol{\omega})
=\bigl(K_\delta(\boldsymbol{\omega})\bigr)^r\). With
\(\kappa(\delta)=\prod_{j=1}^{m}\kappa_j(\delta)\), it obeys
\begin{equation}
 \frac{K_{\beta,r}(\boldsymbol{\omega})}{\kappa(\delta)^r}
 =\left(\prod_j^{\leftarrow}
 e^{-\delta s_j(\boldsymbol{\omega})c_jP_j/2}\right)^r
 \xrightarrow{r\to\infty}e^{-\beta H(\boldsymbol{\omega})/2}.
 \label{eq:app-persistent}
\end{equation}
If ancillas are reused coherently and measured only at the end, the final
record generally hides intermediate histories. Let \(U_t\) denote the
joint system--ancilla unitary at step \(t\), and let \(|\chi_A\rangle\)
be the initial ancilla state. Inserting resolutions of the identity in
the bit-record basis between steps gives the final-outcome operator
\begin{equation}
 \langle\boldsymbol{\omega}_r|U_r\cdots U_1|\chi_A\rangle
 =\sum_{\boldsymbol{\omega}_{r-1},\ldots,\boldsymbol{\omega}_1}
 \langle\boldsymbol{\omega}_r|U_r|\boldsymbol{\omega}_{r-1}\rangle\cdots
 \langle\boldsymbol{\omega}_1|U_1|\chi_A\rangle.
 \label{eq:app-hidden-history}
\end{equation}
Conditioning on the final label therefore combines several intermediate sign histories. Measuring the full trajectory makes those histories accessible, but a trajectory whose signs change still describes an ordered sequence of different Hamiltonians. A static-disorder interpretation requires persistent signs or a separate demonstration that the retained trajectory implements the desired fixed Hamiltonian. The one-step construction used for the main result associates each record with a single signed product from the outset.

For a fixed record, standard product-formula estimates make the remaining approximation explicit~\cite{trotter1959product,suzuki1976generalized,childs2021theory}. Define
\begin{equation}
 \Lambda=\sum_{j=1}^{m}|c_j|,\qquad
 \mathcal C_{\mathrm{comm}}=
 \sum_{j<k}|c_jc_k|\,\|[P_j,P_k]\|.
\end{equation}
The first-order error obeys the uniform bound
\begin{equation}
 \sup_{\boldsymbol{\omega}\in\boldsymbol{\Omega}}
 \left\|\left(\prod_j^{\leftarrow}
 e^{-\beta s_j(\boldsymbol{\omega})c_jP_j/(2r)}\right)^r
 -e^{-\beta H(\boldsymbol{\omega})/2}\right\|
 \leq\frac{\beta^2}{8r}e^{\beta\Lambda/2}\mathcal C_{\mathrm{comm}}.
 \label{eq:app-product-bound}
\end{equation}
The signs do not change the bound because they do not change commutator norms.

\section{Gauge classification and covariant observables}
\label{app:gauge}

Cycle fluxes completely determine the gauge class of bond signs on a connected graph. A spanning-tree construction establishes this classification and provides the local transformations relating any two representatives. Applying those transformations to the measured observables then permits records within a class to be combined.

\subsection{Classification on a connected graph}

After fixing all field signs to \(+1\), a product of local \(X_v\)
conjugations acts on a bond \(e=\{u,v\}\) as
\(s_{\ell(u,v)}(\boldsymbol{\omega})\mapsto
g_u s_{\ell(u,v)}(\boldsymbol{\omega})g_v\), where \(g_v=-1\) if \(X_v\) is
applied and \(g_v=+1\) otherwise. Every vertex on a closed cycle appears
twice, so the cycle product \(\Phi(\gamma,\boldsymbol{\omega})\) is invariant.

\begin{theorem}[Cycle-flux classification]
The bond signs of two records \(\boldsymbol{\omega}\) and \(\boldsymbol{\varpi}\) on a connected
graph are related by local \(X_v\) conjugations if and only if
\(\Phi(\gamma,\boldsymbol{\omega})=\Phi(\gamma,\boldsymbol{\varpi})\) on every independent cycle.
\end{theorem}

\begin{proof}
Necessity follows from the pairwise cancellation of the vertex factors.
For sufficiency, define the relative bond signs
\(r_e=s_{\ell(u,v)}(\boldsymbol{\varpi})s_{\ell(u,v)}(\boldsymbol{\omega})\) for
\(e=\{u,v\}\). Equal cycle products imply
\(\prod_{e\in\gamma}r_e=1\) for every cycle.
Choose a spanning tree and a root \(v_0\), set \(g_{v_0}=1\), and
define \(g_v\) as the product of \(r_e\) along the unique tree path
from \(v_0\) to \(v\). Then \(r_e=g_ug_v\) on every tree edge.
Each non-tree edge closes a fundamental cycle, whose unit product enforces
the same relation on that edge. Hence
\(s_{\ell(u,v)}(\boldsymbol{\varpi})=g_us_{\ell(u,v)}(\boldsymbol{\omega})g_v\) everywhere.
\end{proof}

The cycle-space dimension is
\begin{equation}
 b_1=|\mathcal E|-|\mathcal V|+1,
\end{equation}
so a connected graph has \(2^{b_1}\) bond-sign classes. For an open
\((a+1)\times(b+1)\) square lattice of sites, \(b_1=ab\), and the
elementary plaquettes form an independent cycle basis. The open
\(3\times3\) lattice therefore has four independent plaquette fluxes
and sixteen classes.

Local \(Z_v\) conjugations change the field signs independently without
changing any bond sign. Thus each full-record class \([\boldsymbol{\omega}]\)
includes every field-bit string for each of its bond configurations.
The sum \(\Pr_\beta[\boldsymbol{\omega}]=\sum_{\boldsymbol{\varpi}\in[\boldsymbol{\omega}]}
\Pr_\beta\{\boldsymbol{\varpi}\}\) already includes these field records and must
not be multiplied by an additional \(2^n\). The same applies to the
prescribed class sum. Its full-record prior already contains the
normalizing factor \(2^{-n}\) for the uniformly extended field bits.

\subsection{Pooling observables within a class}

Choose a representative record \(\boldsymbol{\varpi}\in[\boldsymbol{\omega}]\), with positive
field signs. Let \(U\) be a product of local \(X_v\) and \(Z_v\)
operators mapping its signed Hamiltonian to that of \(\boldsymbol{\omega}\), so
\(H(\boldsymbol{\omega})=UH(\boldsymbol{\varpi})U^\dagger\).
For the signed Ising model, the same conjugation transforms each
elementary factor in the fixed-order circuit, so
\(K_\beta(\boldsymbol{\omega})=UK_\beta(\boldsymbol{\varpi})U^\dagger\).
Together with the invariant maximally mixed input, this gives
\begin{equation}
 \rho_\beta(\boldsymbol{\omega})=U\rho_\beta(\boldsymbol{\varpi})U^\dagger,
 \qquad \Pr_\beta\{\boldsymbol{\omega}\}=\Pr_\beta\{\boldsymbol{\varpi}\}.
 \label{eq:app-state-covariance}
\end{equation}
This is an exact covariance of the conditional circuit, independent of
its approximation to a Gibbs state. For an observable \(O(\boldsymbol{\varpi})\)
in the representative frame, define the corresponding observable by
\begin{equation}
 \begin{aligned}
 O(\boldsymbol{\omega})&=UO(\boldsymbol{\varpi})U^\dagger,\\
 \operatorname{Tr}O(\boldsymbol{\omega})\rho_\beta(\boldsymbol{\omega})
 &=\operatorname{Tr}O(\boldsymbol{\varpi})\rho_\beta(\boldsymbol{\varpi}).
 \end{aligned}
 \label{eq:app-observable-covariance}
\end{equation}
For the energy, take \(O(\boldsymbol{\omega})=H(\boldsymbol{\omega})\). If a \(Z\)-basis
measurement gives eigenvalues \(z_v\in\{\pm1\}\), its bond-energy
contribution is
\[
 J\sum_{\{u,v\}\in\mathcal E}s_{\ell(u,v)}(\boldsymbol{\omega})z_uz_v.
\]
If an \(X\)-basis measurement gives \(x_v\in\{\pm1\}\), its
field-energy contribution is
\[
 h\sum_{v\in\mathcal V}s_v(\boldsymbol{\omega})x_v.
\]
Average each contribution over records in the same flux class and then
add the two basis means. The recorded field signs must be included even
though they do not enter the class label. Fixing positive field signs for
the representative therefore does not mean discarding the other field
records. A fixed local observable can be pooled without transformation
only if it is invariant under the relevant local conjugations.

\section{Disorder estimators and sampling uncertainty}
\label{app:estimators}

A prescribed disorder average can be estimated by combining separate class means or by weighting individual shots. Both approaches use the same target prior and require sampling support on its classes. The distribution of records determines their statistical cost and the uncertainty propagated into thermodynamic integration.

\subsection{Stratified and importance-weighted estimators}

Consider \(N_{\mathrm{shots}}\) independent shots drawn from the same
input ensemble. Shot \(r\) produces a complete record \(\boldsymbol{\omega}^{(r)}\)
and a bounded energy estimator \(x^{(r)}\) whose class-conditioned
mean is \(\operatorname{Tr}H(\boldsymbol{\omega})\rho_\beta(\boldsymbol{\omega})\).
For example, choosing \(Z\) or \(X\) measurements with equal
probability and multiplying the corresponding signed energy contribution
in Appendix~\ref{app:gauge} by two gives such an estimator.
Let \(\overline{x}_{[\boldsymbol{\omega}]}\) denote the empirical mean among shots
in class \([\boldsymbol{\omega}]\). Combining the class means gives
\begin{equation}
 E^{\mathrm{shot}}(\beta)
 =\sum_{[\boldsymbol{\omega}]}\Pr[\boldsymbol{\omega}]\,\overline{x}_{[\boldsymbol{\omega}]}.
 \label{eq:app-stratified}
\end{equation}
Every class with nonzero prescribed weight must be sampled to evaluate
this class-mean estimator. In the \(3\times3\) calculation, the two
measurement bases instead have separate fixed budgets. The implemented
estimator is
\[
 E^{\mathrm{shot}}(\beta)=\sum_{[\boldsymbol{\omega}]}\Pr[\boldsymbol{\omega}]
 \left(\overline{x}^{\,Z}_{[\boldsymbol{\omega}]}
       +\overline{x}^{\,X}_{[\boldsymbol{\omega}]}\right),
\]
where the two means contain the signed bond and field contributions,
respectively, without the factor of two used in the randomized-basis
example. For either basis separately, let
\(\Pr_\beta^{\mathrm{shot}}[\boldsymbol{\omega}]\) be its empirical class
frequency. With \(x^{(r)}\) and \(N_{\mathrm{shots}}\) now referring
to that basis, the exact sample closure relation is
\begin{equation}
 \sum_{[\boldsymbol{\omega}]}\Pr_\beta^{\mathrm{shot}}[\boldsymbol{\omega}]
 \overline{x}_{[\boldsymbol{\omega}]}
 =\frac{1}{N_{\mathrm{shots}}}
 \sum_{r=1}^{N_{\mathrm{shots}}}x^{(r)}.
 \label{eq:app-closure}
\end{equation}
The sum runs over sampled classes, since a class with no observations
has no empirical mean. The basis means and frequencies must come from
the same sample for this identity to hold.

An alternative estimator weights individual shots by the ratio of
prescribed to sampling probability. Returning to the independent,
identically distributed full-energy estimator defined above, if
\(\Pr_\beta[\boldsymbol{\omega}]\) is known and is positive wherever
\(\Pr[\boldsymbol{\omega}]>0\), its importance-weighted form is
\begin{equation}
 E^{\mathrm{shot}}(\beta)
 =\frac{1}{N_{\mathrm{shots}}}
 \sum_{r=1}^{N_{\mathrm{shots}}}
 \frac{\Pr[\boldsymbol{\omega}^{(r)}]}{\Pr_\beta[\boldsymbol{\omega}^{(r)}]}
 x^{(r)}.
 \label{eq:app-importance}
\end{equation}
Equations~\eqref{eq:app-stratified} and \eqref{eq:app-importance}
are alternative estimators of \(E^{\mathrm{circ}}(\beta)\), not
identical finite-sample expressions. For the importance-weighted
estimator under these independent, identically distributed sampling
assumptions,
\begin{equation}
 \operatorname{Var}\bigl(E^{\mathrm{shot}}(\beta)\bigr)
 =\frac{1}{N_{\mathrm{shots}}}
 \left(
 \sum_{[\boldsymbol{\omega}]}\frac{\Pr[\boldsymbol{\omega}]^2}{\Pr_\beta[\boldsymbol{\omega}]}
 \mathbb{E}\bigl(x^2\mid[\boldsymbol{\omega}]\bigr)
 -\bigl(E^{\mathrm{circ}}(\beta)\bigr)^2
 \right).
 \label{eq:app-importance-variance}
\end{equation}
The nominal effective sample size,
\begin{equation}
 N_{\mathrm{eff}}
 =\frac{N_{\mathrm{shots}}}
 {\displaystyle\sum_{[\boldsymbol{\omega}]}
 \Pr[\boldsymbol{\omega}]^2/\Pr_\beta[\boldsymbol{\omega}]},
 \label{eq:app-neff}
\end{equation}
measures the cost arising from mismatch between the two distributions.
It is an overlap diagnostic. The actual energy variance also depends on
the conditional second moments in
Eq.~\eqref{eq:app-importance-variance}. Estimating
\(\Pr_\beta[\boldsymbol{\omega}]\) from data introduces additional uncertainty
that must be propagated jointly with the measured energies. The
production run uses balanced sampling of the complete input basis and
separate measurement-basis class means. Its uncertainty is calculated for
that design, rather than by applying the independent-shot formula
unchanged, and is checked against independent sampling repeats in
Appendix~\ref{app:checks}.

\subsection{Reusing class data and quantifying the cost}
\label{app:prior-reuse}

\begin{figure}[htbp]
    \centering
    \includegraphics[width=0.93\linewidth]{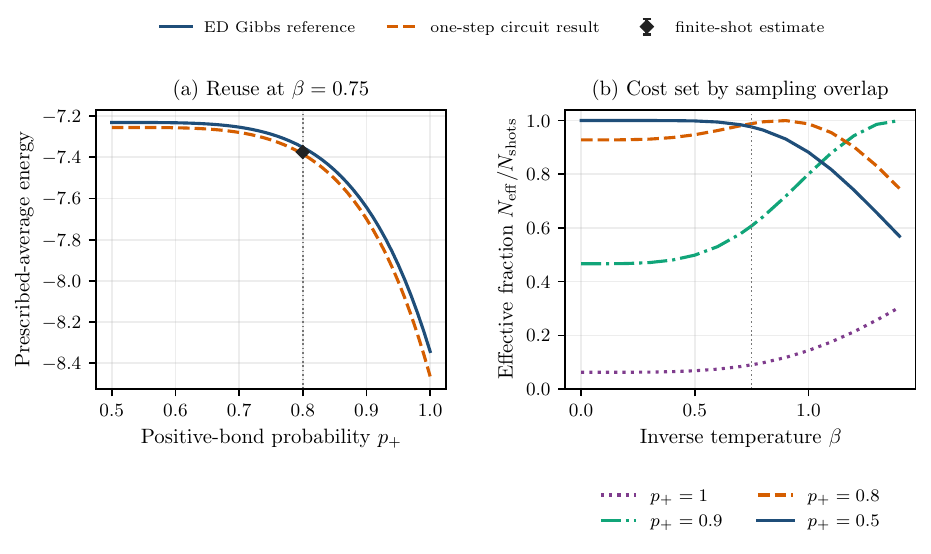}
    \caption{Reuse of resolved class data and its sampling cost.
    (a) The same Gibbs and exact one-step circuit class energies at
    \(\beta=0.75\) give \(E^{\mathrm{diag}}\) and \(E^{\mathrm{circ}}\)
    for each displayed positive-bond probability \(p_+\). The finite-shot
    point is \(E^{\mathrm{shot}}\) at the production choice \(p_+=0.8\),
    with one estimated standard error.
    (b) The nominal effective fraction \(N_{\mathrm{eff}}/N_{\mathrm{shots}}\)
    from Eq.~\eqref{eq:app-neff}, evaluated using \(\Pr[\boldsymbol{\omega}]\) and
    \(\Pr_\beta[\boldsymbol{\omega}]\), for the four fixed priors
    \(p_+=1,0.9,0.8,0.5\). The vertical dotted line marks \(\beta=0.75\).
    This fraction measures distribution overlap. The variance of an energy
    estimate also depends on the class-conditioned measurement fluctuations.}
    \label{fig:prior-reuse}
\end{figure}

The same resolved energies can describe different distributions of
quenched couplings. Figure~\ref{fig:prior-reuse}(a) varies the
positive-bond probability \(p_+\) while holding the class energies at
\(\beta=0.75\) fixed. Panel (b) shows how the nominal effective sample
fraction changes with temperature for several fixed priors. Near
\(\beta=0.75\), the demonstrated \(p_+=0.8\) prior overlaps strongly
with the circuit distribution, whereas \(p_+=1\), which concentrates
the prior on the unfrustrated class, substantially reduces the fraction.

\subsection{Thermodynamic integration}

Integrating the Gibbs identity
\(\langle H(\boldsymbol{\omega})\rangle_\beta
=-\partial_\beta\log\operatorname{Tr}e^{-\beta H(\boldsymbol{\omega})}\) gives
\begin{equation}
 \log\operatorname{Tr}e^{-\beta H(\boldsymbol{\omega})}
 =n\log2-\int_0^\beta
 \langle H(\boldsymbol{\omega})\rangle_\alpha\,\dd\alpha.
 \label{eq:app-logZ}
\end{equation}
The fixed prior permits the class average to pass through the integral,
giving
\[
 F(\beta)=-\frac{n\log2}{\beta}
 +\frac{1}{\beta}\int_0^\beta E(\alpha)\,\dd\alpha.
\]
A temperature-dependent prior would contribute additional terms when
differentiating the averaged log partition function, so the quenched
calculation uses the same \(\Pr[\boldsymbol{\omega}]\) throughout the sweep.
Substituting the circuit energy curve defines
\[
 F^{\mathrm{circ}}(\beta)=-\frac{n\log2}{\beta}
 +\frac{1}{\beta}\int_0^\beta E^{\mathrm{circ}}(\alpha)\,\dd\alpha,
 \qquad
 F^{\mathrm{circ}}(\beta)-F(\beta)
 =\frac{1}{\beta}\int_0^\beta
 \bigl(E^{\mathrm{circ}}(\alpha)-E(\alpha)\bigr)\,\dd\alpha.
\]
Thus the free-energy bias is the integrated circuit-energy bias.
Replacing \(E^{\mathrm{circ}}\) by finite-shot estimates would produce
\(F^{\mathrm{shot}}\). The central circuit curve in
Fig.~\ref{fig:quenched-free-energy} instead uses exact circuit expectations.
Its band propagates projected sampling variances through the Simpson
integration weights, using the sampling prescription of
Appendix~\ref{app:resources} and independent runs at distinct temperatures
for that projection. The ED comparison quantifies finite-step bias,
while refining the temperature grid quantifies quadrature error.
These systematic contributions are reported separately from the projected
sampling band in Appendix~\ref{app:checks}.

\FloatBarrier
\section{Numerical validation and additional results}
\label{app:checks}

Small lattices allow direct comparisons of the conditional operators, record probabilities, and Gibbs targets. These calculations test the map and gauge covariance independently of the prescribed-average reconstruction. Additional checks on the \(3\times3\) data assess integration error, sampling uncertainty, and the spatial information retained by the flux classes.

\subsection{Input-state and resource checks}
\label{app:resources}

The complete-basis construction can be checked against purification on the \(2\times2\) lattice. At \(\beta=0.75\), averaging all sixteen computational-basis inputs reproduces the purified record distribution with total-variation distance \(1.2\times10^{-11}\). Using only the all-zero input gives a distance of \(0.465\), reflecting the dependence of record probabilities on the initial state. This comparison verifies the ensemble preparation used in the larger calculation.

The simulated \(3\times3\) circuit uses nine system qubits and twenty-one record qubits, giving a total width of thirty qubits. Preparing a purification would add nine reference qubits. The complete-basis implementation exchanges that additional register for separate preparations of the \(512\) system basis states. Hardware implementations would need to account for the cost of these preparations together with circuit depth and measurement noise. At \(\beta=0.75\), each basis input is measured with \(128\) shots in each of the \(Z\) and \(X\) bases, giving \(65\,536\) shots per basis. These counts specify \(E^{\mathrm{shot}}(0.75)\) and its reported
standard error. The same per-basis budget sets the projected sampling
uncertainty along the thermodynamic-integration curve, whose central
value \(F^{\mathrm{circ}}\) is calculated from exact circuit
expectations without a separate finite-shot run at every temperature.

The record indices in the signed Hamiltonian label field bits first and
bond bits second. The implementation stores bond bits first and field
bits second. Translating to the manuscript convention is only a
permutation of record labels. It does not change the physical circuit
order, which applies the \(ZZ\) interactions before the \(X\)
interactions. The probabilities, flux labels, and signed energy
contributions are evaluated using the corresponding bond and field bits.

\Needspace{8\baselineskip}
\subsection{Conditional-map and smaller-lattice checks}

On the open \(2\times2\) lattice, all conditional operators can be extracted explicitly. At \(J=h=\pi/4\) and \(\beta=0.75\), optimizing over the overall scalar leaves an elementary-map error of \(1.7\times10^{-16}\) for the exact angle, compared with \(1.57\times10^{-2}\) for the linear-angle approximation. Combining all eight noncommuting terms gives a mean conditional-state trace distance of \(0.1090\) from the ED Gibbs targets. The exact elementary map therefore removes the angle-approximation error, while the full layer retains the product-formula error summarized in Table~\ref{tab:branch-certification}.

\begin{table}[htbp]
    \caption{Conditional-map comparisons on the open \(2\times2\) lattice at \(J=h=\pi/4\) and \(\beta=0.75\). Here \(\overline d\) is the mean trace distance from the Gibbs states of the final-label Hamiltonians. The quantity \(D_{\TV}\) is the total-variation distance between normalized retained-record probabilities and their partition-function weights. Coverage is the total probability of retained records. Persistent fresh records admit a static multi-step interpretation with low coverage. For coherent reuse, the final-label Hamiltonians serve only as comparison targets.}
    \label{tab:branch-certification}
    \centering
    \small
    \setlength{\tabcolsep}{8pt}
    \begin{tabular}{@{}lcccc@{}}
        \toprule
        retained record & static \(H(\boldsymbol{\omega})\)? & \(\overline d\) & \(D_{\TV}\) & coverage \\
        \midrule
        one step, exact angle & one-layer label & 0.1090 & 0.01090 & 1.0000 \\
        one step, linear angle & one-layer label & 0.1337 & 0.01932 & 1.0000 \\
        two steps, coherent reuse & no & 0.0918 & 0.01330 & 1.0000 \\
        two steps, persistent fresh records & yes & 0.0510 & 0.00259 & 0.00675 \\
        \bottomrule
    \end{tabular}
\end{table}

Gauge covariance remains accurate even when the one-step circuit differs from the target Gibbs state. On the \(2\times2\) lattice, transforming every conditional state to a common representative frame leaves a maximum within-class trace distance of \(1.6\times10^{-16}\) and a maximum relative probability difference of \(1.2\times10^{-16}\). On the \(2\times3\) lattice, gauge-related representatives have spectra within \(2.4\times10^{-14}\), thermal energies within \(1.5\times10^{-14}\), and log partition functions within \(6.3\times10^{-15}\).

\FloatBarrier
\begin{figure}[htbp]
    \centering
    \subfloat[]{\includegraphics[width=0.48\linewidth]{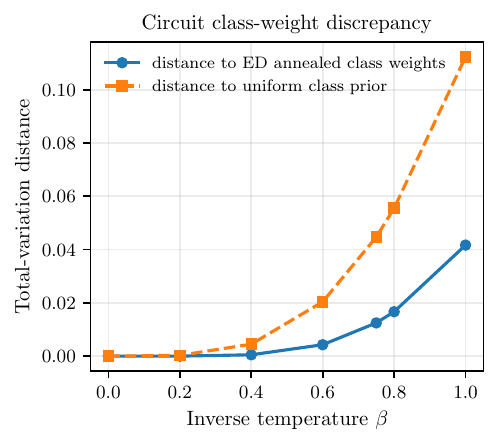}}
    \hfill
    \subfloat[]{\includegraphics[width=0.48\linewidth]{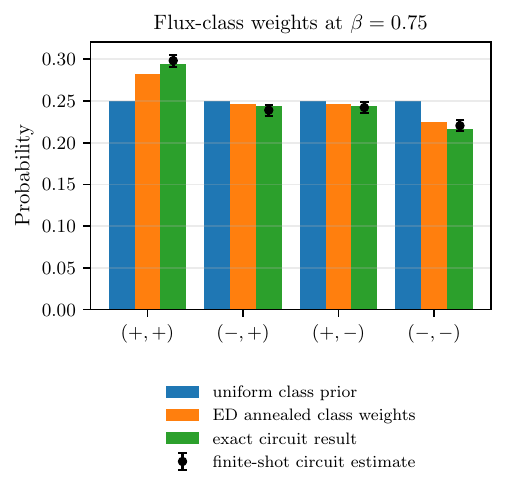}}
    \caption{Circuit sampling probabilities on the open \(2\times3\) lattice.
    (a) Total-variation distances between the exact circuit class probabilities
    \(\Pr_\beta[\boldsymbol{\omega}]\) and, respectively, the ED extended-Gibbs class
    weights and the uniform class prior. The extended-Gibbs weights are
    \(\sum_{\boldsymbol{\varpi}\in[\boldsymbol{\omega}]}\operatorname{Tr}e^{-\beta H(\boldsymbol{\varpi})}
    /\sum_{\boldsymbol{\varpi}\in\boldsymbol{\Omega}}\operatorname{Tr}e^{-\beta H(\boldsymbol{\varpi})}\).
    For the uniform microscopic prior used in this comparison, they are the
    annealed class weights.
    (b) The three distributions are shown at \(\beta=0.75\). Black points are the
    \(4096\)-shot circuit estimates with one-standard-error bars. Each sign
    pair specifies the two plaquette fluxes and hence a class \([\boldsymbol{\omega}]\).
    The comparisons distinguish the circuit's measured distribution from both
    equilibrium and uniform references. They do not define the prescribed
    \(p_+=0.8\) prior of the \(3\times3\) calculation.}
    \label{fig:small-class-distribution}
\end{figure}

\begin{figure}[htbp]
    \centering
    \subfloat[]{\includegraphics[width=0.48\linewidth]{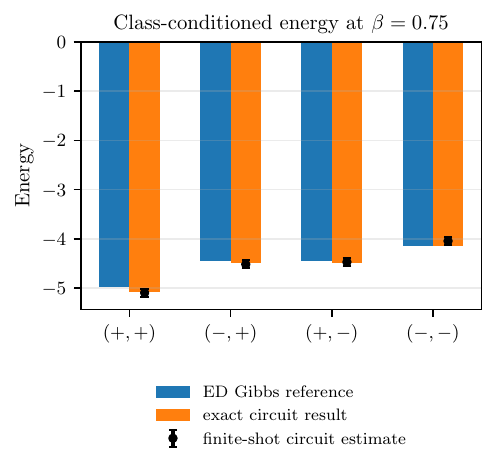}}
    \hfill
    \subfloat[]{\includegraphics[width=0.48\linewidth]{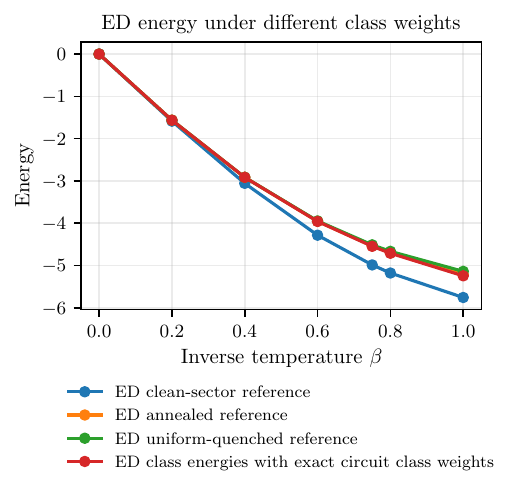}}
    \caption{Class conditioning on the open \(2\times3\) lattice.
    (a) Gibbs expectations \(\langle H(\boldsymbol{\omega})\rangle_\beta\) from ED,
    exact one-step circuit expectations
    \(\operatorname{Tr}H(\boldsymbol{\omega})\rho_\beta(\boldsymbol{\omega})\), and
    \(4096\)-shot estimates at \(\beta=0.75\). Sign pairs identify the
    plaquette-flux classes.
    (b) The same Gibbs class energies combined with a prior concentrated on
    the unfrustrated class, the uniform-prior annealed weights, a uniform
    quenched class prior, or the exact circuit probabilities
    \(\Pr_\beta[\boldsymbol{\omega}]\). The last curve is
    \(\sum_{[\boldsymbol{\omega}]}\Pr_\beta[\boldsymbol{\omega}]\langle H(\boldsymbol{\omega})\rangle_\beta\).
    This curve changes only the weights and is not
    \(\widetilde E^{\mathrm{circ}}\), which also uses circuit class energies.
    The separation between the curves isolates the effect of the weighting.}
    \label{fig:small-energy-reconstruction}
\end{figure}

\FloatBarrier
\subsection{Grid and finite-shot checks}

The temperature-grid comparison quantifies the quadrature contribution
to the endpoint free energy. At \(\beta=1.4\), integration of the ED
class energies on fifteen points differs from \(F^{\mathrm{diag}}\)
by \(-2.20\times10^{-5}\). Refining the circuit integration from
fifteen to twenty-nine points changes the numerical estimate of
\(F^{\mathrm{circ}}\) by \(-1.84\times10^{-5}\). Both changes are
small compared with the finite-step bias
\(F^{\mathrm{circ}}-F^{\mathrm{diag}}=+0.176702\).
The separate ED comparison gives a quenched-minus-annealed free-energy
difference of \(0.107066\) for the same prior.

The finite-shot energy \(E^{\mathrm{shot}}(0.75)\) reported in the
main text uses base seed \(2026\). The sampling check comprises twelve
runs, including this realization, with non-overlapping base-seed blocks
and the same measurement budget. Their mean, \(-7.386246\), agrees with the exact one-step circuit expectation \(-7.384441\). The empirical standard deviation is \(0.009381\), compared with a mean estimated standard error of \(0.013576\). The largest residual divided by its estimated standard error is \(1.412\) for the prescribed average and \(2.665\) for an individual class. The maximum error in the reconstruction closure relation, Eq.~\eqref{eq:app-closure}, is \(8.88\times10^{-16}\). These repeats support the reported uncertainty scale, with the precision of the comparison limited by the number of repeats.

\FloatBarrier
\subsection{Additional spatial information}

\begin{figure}[htbp]
    \centering
    \includegraphics[width=0.84\linewidth]{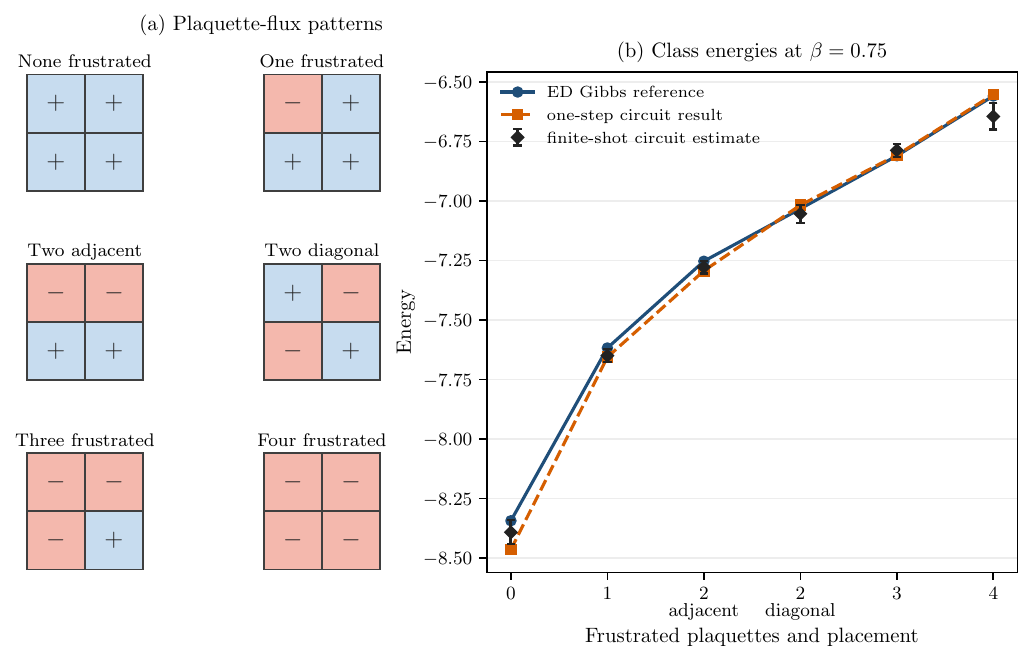}
    \caption{Additional spatial information in the open \(3\times3\) class data.
    (a) Representative plaquette-flux patterns for the six groups related by
    rotations and reflections. Blue and red denote \(\Phi=+1\) and
    \(\Phi=-1\), respectively. Each full pattern identifies a flux class
    \([\boldsymbol{\omega}]\).
    (b) Gibbs class energies from ED, exact one-step circuit expectations,
    and finite-shot estimates at \(\beta=0.75\), with the latter pooled
    within each symmetry group and shown with one estimated standard error.
    The diagonal-minus-adjacent energy difference for two frustrated
    plaquettes is \(+0.220213\) for the Gibbs target and \(+0.275646\)
    for the exact circuit calculation. Equal numbers of frustrated
    plaquettes therefore need not imply equal class energies.}
    \label{fig:spatial-resolution}
\end{figure}

The \(3\times3\) class energies distinguish the spatial arrangement of frustration. With two frustrated plaquettes, placing them next to one another or across a diagonal produces different energies even though the number of negative fluxes is unchanged. Figure~\ref{fig:spatial-resolution} displays the representative patterns and their energy splitting, illustrating the information retained by conditioning on the full flux class.

\FloatBarrier

\end{document}